# Quantum-Annealing Enhanced Machine Learning for Interpretable Phase Classification of High-Entropy Alloys

Diego Ibarra Hoyos[1]*, Gia-Wei Chern[1], Israel Klich[1], Joseph Poon[1,2,*]

[1] Department of Physics, University of Virginia, Charlottesville, VA 22904

[2] Department of Materials Science and Engineering, University of Virginia, Charlottesville, VA 22904

*Corresponding author. Email: sjp9x@virginia.edu (J.P.),di8pd@virginia.edu (D.I.H.)

**ABSTRACT**

High-entropy alloys (HEAs) offer unprecedented compositional flexibility for designing advanced materials, yet predicting their crystallographic phases remains a key bottleneck due to limited data and complex phase formation behavior. Here, we present a quantum-enhanced machine learning framework that leverages quantum annealing to enhance phase classification in HEAs. Our pipeline integrates Quantum Boosting (QBoost) for interpretable feature selection and classification, with Quantum Support Vector Machines (QSVM) that use quantum-enhanced kernels to capture nonlinear relationships between physical descriptors. By reformulating both models as Quadratic Unconstrained Binary Optimization (QUBO) problems, we exploit the efficient sampling capabilities of quantum annealers to achieve rapid training and robust generalization, demonstrating notable runtime reductions relative to classical baselines in our setup. We target six key phases: FCC, BCC, Sigma, Laves, Heusler, and Al–X–Y B2, and benchmark model performance using both cross-validation and a rigorously curated test set of prior experimentally synthesized HEAs. The results confirm strong alignment between predicted and measured phases. Our findings demonstrate that quantum-enhanced classifiers match or exceed classical models in accuracy and offer insights grounded in interpretable physical descriptors. This work constitutes an important step toward practical quantum acceleration in materials discovery pipelines.

## INTRODUCTION

High Entropy Alloys (HEAs) are a class of materials composed of multiple principal elements, typically four or more, in near-equiatomic proportions**.** This compositional strategy introduces vast chemical and configurational freedom, enabling broad design of properties for diverse applications, from high-temperature structural components to corrosion-resistant coatings, as well as functional applications such as magnetic, electrical, and catalytic systems **[1-10]** As the number of constituent elements increases, the design space grows exponentially with the variations in compositions **[11]**.

A central challenge in HEA design lies in predicting the resulting phase, typically solid solution (SS), intermetallic (IM), or a combination (SS + IM), as the phase strongly governs the alloy's physical behavior. However, the vastness of the compositional space makes experimental or trial-and-error approaches infeasible at scale. This has led to the adoption of machine learning models for phase classification, enabling faster screening and design of candidate alloys**[12-21]**. Yet, conventional machine learning techniques often struggle with the limited size and imbalance of materials datasets, the complex high-dimensional relationships between composition and phase, and the significant computational cost associated with feature engineering and model optimization; and issue that becomes especially pronounced in iterative frameworks like active learning, where repeated retraining amplifies these burdens**[13,17, 20]**.

In parallel, quantum computing, particularly quantum annealing (QA), has emerged as a novel strategy to accelerate materials discovery**[21-27]**. By leveraging quantum mechanics to solve combinatorial optimization problems, quantum annealing can enhance classical machine learning **[28-32]**. Specifically, quantum-boosted (QBoost) methods **[28,30-32]** and Quantum Support Vector Machines (QSVM) **[29]** seem to offer higher expressibility and better generalization, especially in low-data regimes**[33,34]**. These approaches have been explored in domains ranging from particle physics and software validation to fluid dynamics and molecular property prediction **[28–34,35].** While hybrid quantum–classical models have recently been applied to HEA phase classification using variational quantum circuits on gate-based simulators **[32],** our work is the first to leverage quantum annealing–based QBoost and QSVM for both feature selection and classification, integrating physically informed descriptors into a modular and interpretable framework.

Previous work has demonstrated that quantum and classical annealing-based boosting algorithms deliver classification accuracy competitive with deep learning models like Deep Neural Networks (DNNs) and Extreme Gradient Boosting (XGBoost), while maintaining robustness to label noise and overfitting **[28,33,34]**. Importantly, their quantum annealing-based model yielded interpretable classifiers grounded in physically meaningful features, an essential attribute in materials science, where datasets are small and explainability matters**[36]**.

In the present work, we explore quantum-driven algorithms, namely QBoost and QSVM, for phase identification in HEAs. Specifically, we aim to classify key crystallographic phases commonly observed in HEAs, including face-centered cubic (FCC), body-centered cubic (BCC), refractory B2 (RB2), Heusler, Sigma, and Laves phases. By formulating phase classification as a Quadratic Unconstrained Binary Optimization (QUBO) problem and solving it via quantum annealing (QA), our approach efficiently navigates limited data while delivering physically interpretable predictions. In addition, by integrating quantum annealing into the training pipeline, we observe significant reduction in computational costs in specific stages of model optimization, though this remains hardware- and problem-dependent. To validate our models, we benchmark their performance not only through cross-validation but also against an independent test set of HEAs that we experimentally synthesized and characterized in a prior study[**20**]. This quantum-enhanced pipeline offers a practically significant path toward accelerating HEA design through principled, data-driven discovery

## RESULTS

### Quantum-enchanced Machine Learning Framework

Accurate phase identification in HEAs remains a core challenge due to their compositional complexity and the computational demands of navigating a vast, high-dimensional feature space. While classical machine learning methods have progressed **[13, 17, 20]**, they can face challenges such as limited interpretability, reduced performance in data-scarce regimes, and increased computational demands when extensive feature engineering or model tuning is required.

To overcome these limitations, we developed a quantum-enhanced machine learning framework designed to improve accuracy and interpretability while substantially reducing computational cost. Building on insights from prior work **[20],** which used logistic regression with exhaustive feature expansion and sequential feature selection, we aimed to address broader challenges in materials informatics, namely, generalization in low-data regimes, and the computational burden of feature optimization. The proposed pipeline strategically integrates QBoost and QSVM, leveraging QA for efficient optimization.

Quantum annealing, used throughout the pipeline, explores complex energy landscapes via quantum tunneling, solving the underlying QUBO problems with high efficiency **[37]**. Unlike gate-based quantum computers, which implement general-purpose logic through sequences of quantum gates, quantum annealers are application-specific devices tailored for optimization and are already available as commercial hardware. This allows the proposed quantum-enhanced pipeline to capitalize on near-term quantum resources while remaining interpretable, adaptive, and computationally tractable.

As illustrated in Fig. 1, the pipeline begins with the construction of a physics-informed feature set comprising 31 descriptors derived from properties of elements and binary alloys. These include features based on phase diagram heuristics **[18]**, Hume-Rothery rules**[38]**, and other thermodynamic and electronic principles known to govern phase formation in multicomponent systems (Table 1). A complete list and description of these features is provided in Supplementary Table 1.

This feature set (Table 1) is then passed to a QBoost model, which performs both classification and feature selection. QBoost constructs an ensemble of weak classifiers, each trained on a single feature, and uses QA to determine the optimal subset to include in the final model. Because weak classifiers are constrained to one feature each, the inclusion or exclusion of individual models directly maps to feature relevance. This enables QBoost to serve not only as an interpretable classifier, but also as a physics-aware feature selector. Importantly, this interpretability stems from the boosting framework itself, specifically, the use of single-feature decision stumps and binary weights, while quantum annealing contributes primarily to optimization efficiency and potential accuracy gains.

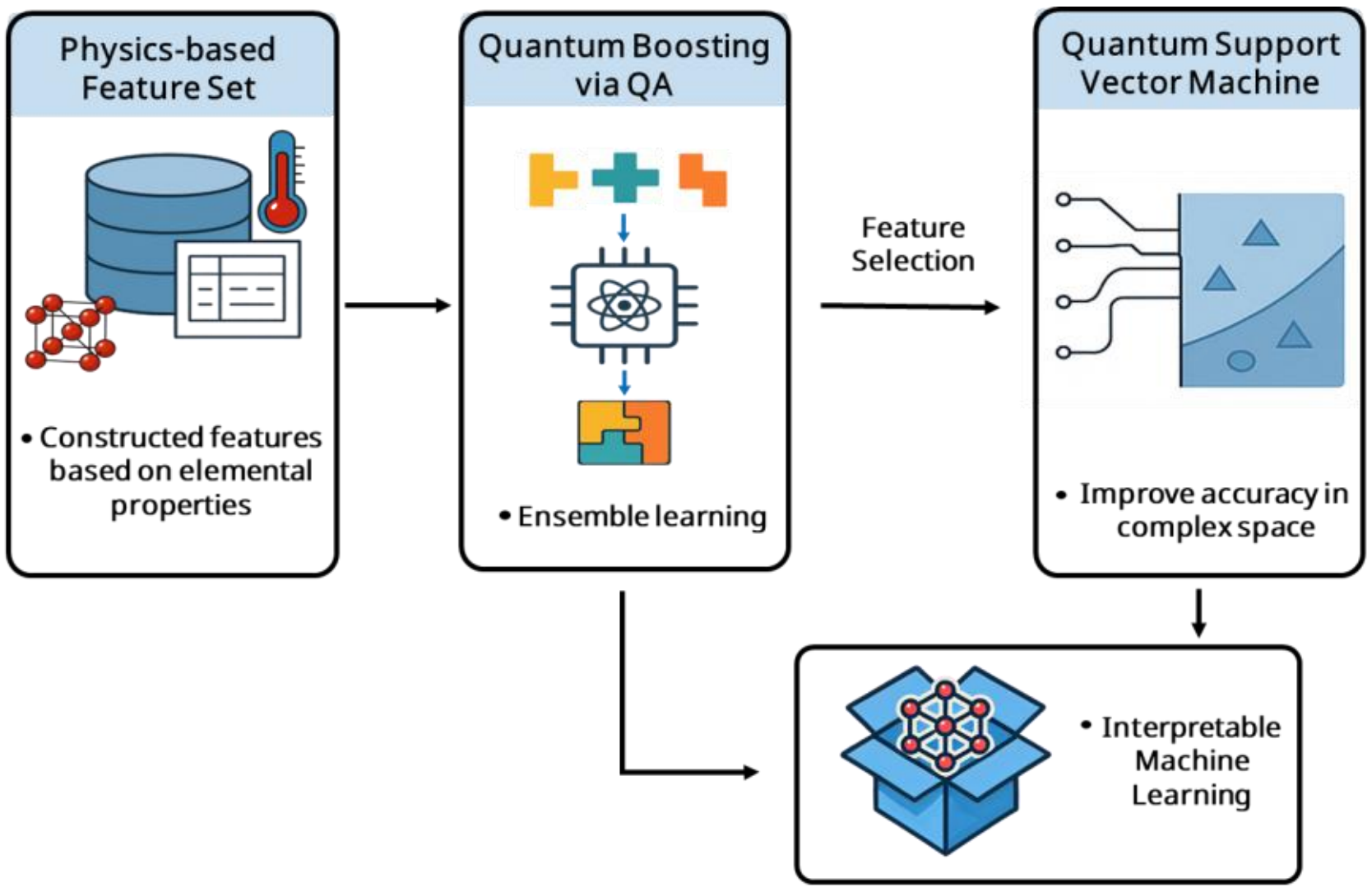

**Figure 1. Schematic overview of the quantum-enhanced machine learning pipeline for HEA phase classification.** The workflow begins with the construction of a physics-informed feature set comprising 31 descriptors based on elemental properties and thermodynamic principles. These features are then passed to a Quantum Boosting model, performing initial classification. If Quantum Boosting performance is insufficient, the selected features are passed to a Quantum Support Vector Machine to capture higher-order interactions via quantum-enhanced kernels. The pipeline provides an adaptive, interpretable, and computationally efficient strategy for phase prediction in high-entropy alloys.

To balance efficiency and accuracy, the pipeline is adaptive: if QBoost alone achieves satisfactory performance, its compact and transparent model is retained for final classification. However, in cases where QBoost reaches a performance ceiling, particularly in distinguishing complex intermetallic phases, we repurpose QBoost from a final classifier to a feature selection stage, using the features it selects as input to a QSVM. Although not explicitly optimized for QSVM, this subset represents a high-performing, interpretable feature set that enables efficient learning while preserving consistency across the pipeline. This second stage uses quantum-enhanced kernel methods to model higher-order, nonlinear interactions that QBoost may underfit, thus improving classification accuracy while maintaining the physically grounded feature space.

**Table 1 | Subset of thermodynamic and Hume-Rothery rule features used in this study.**

A subset of the complete list of features used in the Machine Learning model [**20**]. The full set can be found in Supplementary Table 1.

| **Formula** | **Comments** |
|---|---|
| Mixing Entropy: $\Delta S_{mix} = -R\sum_{i=1}^{N} c_i \ln(c_i)$ | R: The gas constant. $c_i$: The atomic percentage of the i-th element for a N-component system. (Definitions of N and $c_i$ are the same elsewhere.) |
| Mixing Enthalpy: $\Delta H_{mix} = \sum_{i=1,i\neq j}^{N} 4\,\Delta H_{i,j}^{mix}\, c_i c_j$ | $\Delta H_{i,j}^{mix}$: The binary mixing enthalpy obtained from Miedema's model of i-j element pair. |
| Tmelt: $T_m = \frac{\sum_{i\neq j} T_{i-j}\times c_i \times c_j}{\sum_{i\neq j} c_i \times c_j}$ | $T_{i-j}$ is the melting temperature of the i-j elements for the relative ratio of the two elemental concentrations $c_i$ and $c_j$ of the HEA composition. |
| Valence Electron Concentration: $VEC = \sum_{i=1}^{N} c_i\, VEC_i$ | $VEC_i$: Valence electrons count of the i-th element. |
| Omega: $\Omega = \frac{T_m \Delta S_{mix}}{\lvert\Delta H_{mix}\rvert}$ | |
| Phi2: $\Phi = \frac{\Delta G_{SS}}{-\lvert G_{max}\rvert}$ | $\Delta G_{SS}$: The Gibbs free energy change for forming a fully disordered SS phase. $\Delta G_{max}$: The larger absolute Gibbs free energy change of forming the strongest binary compound, or having phase segregation. |
| Eta: $\eta = \frac{-T_{ann}\Delta S_{mix}}{\lvert\Delta H_f\rvert}$ | $T_{ann}$: Annealing temperature. If $T_{ann}$ is not known, use $T_{ann} = 0.8\ T_m$. $\Delta H_f$: The most negative binary mixing enthalpy for forming IM |
| $\frac{E_2}{E_0} = \sum_{j\geq i}^{N} \frac{c_i c_j \lvert r_i + r_j - 2\bar{r}\rvert^2}{(2\bar{r})^2}$ | $\bar{r} = \sum_{i=1}^{N} c_i\, r_i$: Average atomic radius. |

## Quantum Boosting

Boosting is an ensemble learning method that constructs a robust classifier

$$H(x) = sign\left(\sum_{i=1}^{N} \omega_i h_i(x)\right) \tag{1}$$

by combining multiple weak classifiers $\{h_i\}_{i=1}^{N}$, each defined as $h_i: \mathbb{R}^M \to \{-1, +1\}$, into an ensemble that achieves robust classification performance[**31**]. Here, $M$ denotes the feature dimensionality and $N$ the total number of weak classifiers. A weak classifier is an algorithm whose individual performance is marginally better than random guessing, and in our implementation, each $h_i$ is trained as a decision stump on a single input feature, allowing us to isolate the contribution of each physical descriptor and improve interpretability. This design is especially suited for applications like phase prediction in HEAs, where understanding the role of individual features is as important as classification accuracy.

To implement boosting within a quantum framework, we minimize a loss function balancing classification error with sparsity by optimizing the binary weight vector $\omega$.

$$\min_{\omega\in\{0,1\}^N}\left[\sum_{s=1}^{S}\left(y_s-\sum_{i=1}^{N}\omega_i h_i(x_s)\right)^2+\lambda\|\omega\|_0\right] \tag{2}$$

Where $\{(x_s, y_s)\}_{s=1}^{S}$ are the $S$ training samples with binary labels $y_s \in \{-1,+1\}$, and $\|\omega\|_0$ count the number of active classifiers. The regularization parameter $\lambda$ controls the trade-off between sparsity and classification fidelity.

We adopt a squared loss function, rather than conventional exponential or hinge losses, to directly translate the optimization problem into a QUBO, compatible with quantum annealers. This form is naturally compatible with quantum annealing hardware, such as D-Wave systems, which solve Ising-type Hamiltonians. In this formulation, the binary weights $\omega_i \in \{0,1\}$ become spin variables, and quantum annealing searches for the optimal spin configuration minimizing the Ising energy landscape.

This approach, known as QBoost, diverges from classical boosting (e.g., AdaBoost **[39]**) in two key ways: (1) the weights are binary, yielding more interpretable models with clear feature inclusion/exclusion, and (2) $L_0$-norm regularization directly enforces sparsity, allowing the ensemble to perform well with fewer active classifiers**[30,31]**. Each selected weak learner maps to a single physical descriptor, producing compact, transparent models with traceable mechanisms.

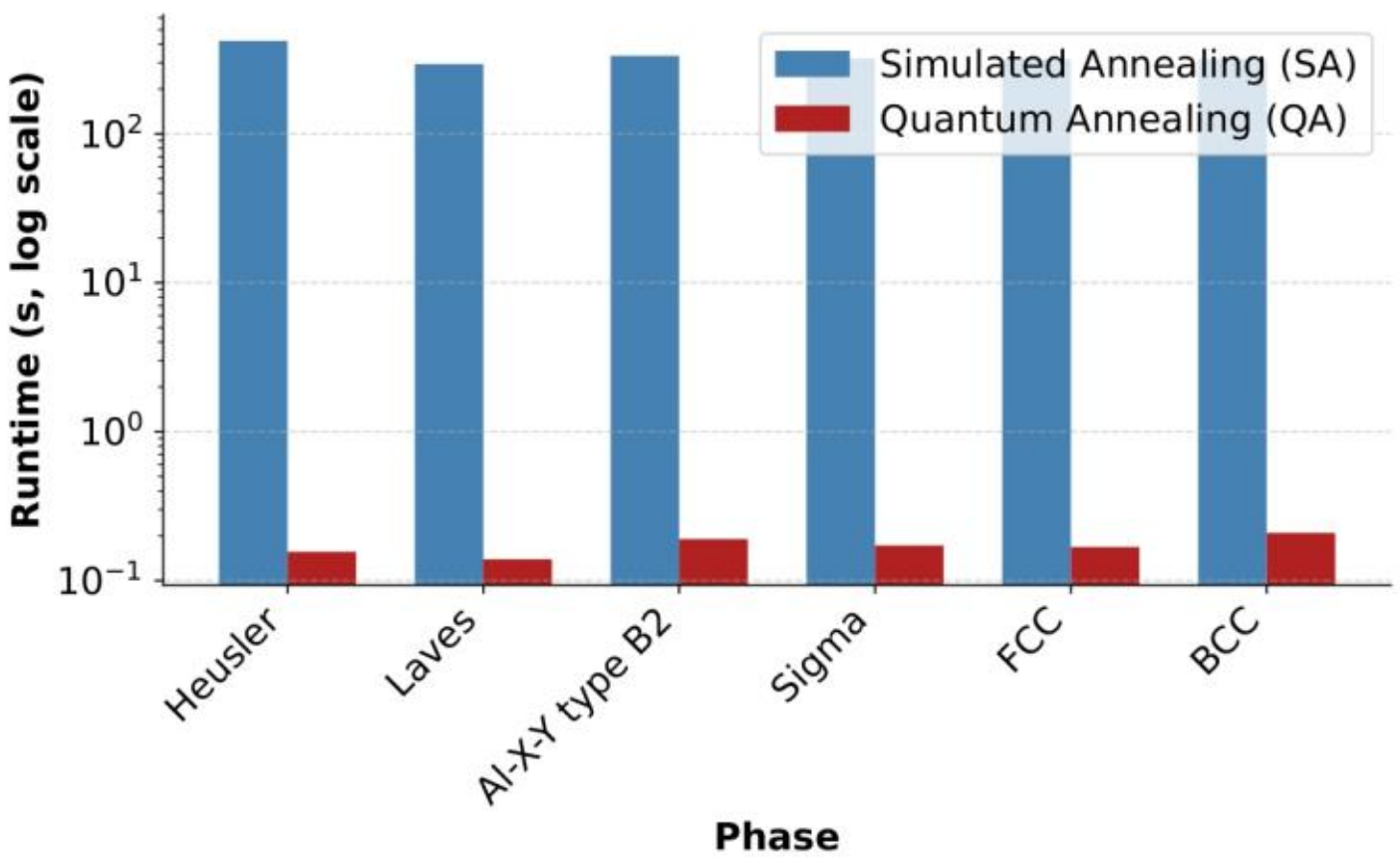

**Fig 2. Quantum vs. Simulated Annealing runtime comparison.** Bar plots comparing total runtime (log scale) for weak classifiers using SA (blue, executed on UVA's Rivanna/Afton node) and QA (red, executed on D-Wave quantum annealer) across six structural phases. QA achieves orders-of-magnitude faster convergence.

We applied QBoost to classify six structural phases in HEAs (FCC, BCC, B2+, Heusler+, Sigma+, and Laves+) and benchmarked performance against simulated annealing (SA). Quantum annealing, executed using the D-Wave Advantage quantum annealer, solved each QUBO instance in approximately 0.01–0.02 seconds, representing a roughly 10,000x acceleration over simulated annealing (~300–400 seconds per instance on UVA's Rivanna High-Performance Computing Cluster, equipped with Intel Xeon Platinum CPUs) (Fig. 2). While this practical acceleration significantly reduces computational time from hours to seconds, it primarily results from quantum tunneling efficiencies and hardware-specific advantages rather than demonstrably superior computational complexity scaling. Therefore, rigorous complexity analysis remains essential to fully characterize quantum annealing’s scalability in a theoretical computational context. The full datasets and 5-fold cross-validation benchmarks are detailed in Methods

In terms of accuracy, quantum annealing consistently outperformed or matched SA across all datasets. Table 2 summarizes the results. On the testing sets, composed of unseen alloys that we cast and characterized in our lab using X-ray diffraction, QA surpassed SA in every case except for the RB2 phase, where both methods performed equally. These gains in performance and efficiency highlight QA’s advantage in both model quality and runtime.

**Table 2 | Classification performance of SA- and QA-based models across structural phases.** Accuracy and macro-averaged F1 scores are reported for models trained via SA and QA feature selection. Performance is evaluated on both training (5-fold cross-validation) and testing (unseen experimentally synthesized alloys). QA-selected models consistently achieve comparable or superior generalization with significantly reduced computational cost.

| Phase | Accuracy (Train) | | F1-score (Train) | | Accuracy (Test) | | F1-score (Test) | |
|---|---|---|---|---|---|---|---|---|
| | QA | SA | QA | SA | QA | SA | QA | SA |
| FCC | 0.9165 | 0.9052 | 0.9160 | 0.9046 | **0.8833** | 0.8500 | 0.8919 | 0.8627 |
| BCC | 0.9410 | 0.9410 | 0.9408 | 0.8168 | **0.9333** | 0.9333 | 0.9313 | 0.9313 |
| RB2 | 0.7549 | 0.6856 | 0.7576 | 0.6011 | **0.7143** | 0.7143 | 0.6976 | 0.6450 |
| Heusler | 0.7866 | 0.7806 | 0.7738 | 0.7434 | **0.5833** | 0.5000 | 0.55 | 0.4375 |
| Laves | 0.8228 | 0.7862 | 0.8208 | 0.5917 | **0.8667** | 0.8000 | 0.8978 | 0.8278 |
| Sigma | 0.8567 | 0.8181 | 0.8451 | 0.6530 | **0.6333** | 0.5333 | 0.7230 | 0.6433 |

Cubic phases (FCC and BCC) were modeled particularly well, likely due to their dependence on well-understood descriptors such as valence electron concentration (VEC) **[40-42]**. In contrast, complex intermetallic phases, RB2, Laves, and Heusler, posed greater challenges due to their multivariate dependencies, nonlinear electronic effects, and long-range interactions **[43,44]**. Nonetheless, even when predictive accuracy declined, feature selection results aligned with known structure, property relationships and remained valuable for downstream analysis. This interpretability stems from the nature of the QBoost model itself: by using decision stumps on raw, physics-informed descriptors, without transforming them into abstract feature spaces, QBoost maintains the original scientific significance of each feature. This enables direct attribution of selected features to well-established mechanisms governing phase stability.

To illustrate the physical insights enabled by our quantum-enhanced feature selection, we constructed a feature–phase matrix, mapping individual physical descriptors against different HEA phase types (Fig. 3). For FCC and BCC phases, key descriptors such as VEC, PFP_A1, and PFP_A2 were consistently selected, reflecting their established crystallographic relevance **[18,20,40-42]**. While feature patterns for complex intermetallics like Laves, Sigma or Heusler were less consistent, stability-related descriptors $\Omega$ and $\eta$ appeared across multiple phases, underscoring their relevance in predicting phase stability. **[20,45,46]**.

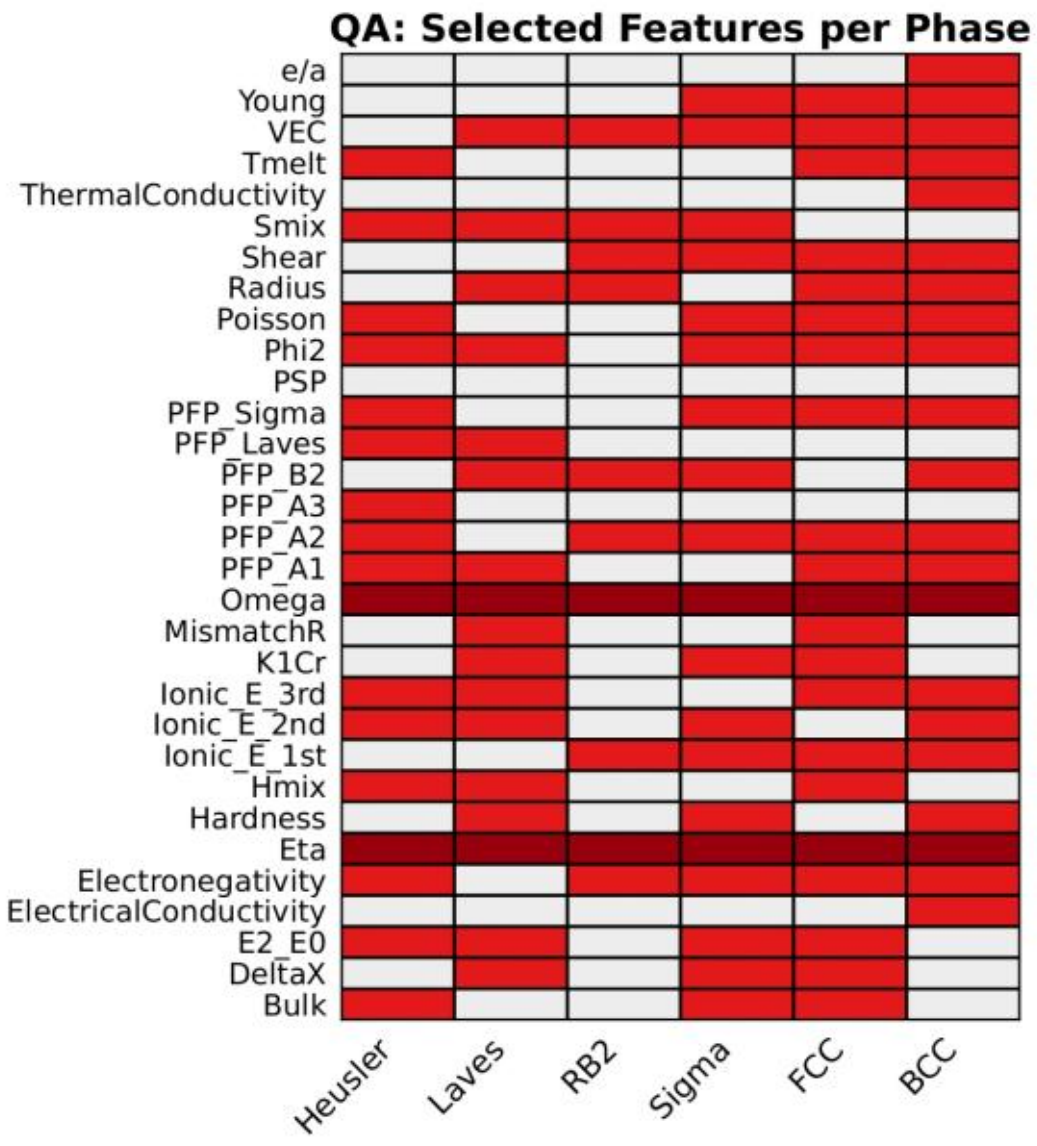

**Fig 3. Feature selection heatmap by Quantum Boosting (QBoost).** Red indicates selected features; white indicates excluded features, highlighting phase-specific descriptor relevance across different crystal structures.

To further examine the interpretability and discriminative power of the quantum-selected descriptors, we present detailed feature histograms for BCC, the phase where QBoost achieved the highest predictive accuracy (**Fig. 4a**), and for Heusler, the phase with the lowest predictive accuracy (**Fig. 4b**). For BCC, we contrast the distributions of selected features (VEC, PFP_A2, Tmelt) with a non-selected feature (Smix). The selected descriptors display clear class separability, underscoring their relevance, while Smix shows substantial overlap between classes, consistent with its exclusion by QBoost. In contrast, the Heusler phase exhibits no evident separability even among selected features (Omega, Phi2, E2_E0), suggesting that classification in this intermetallic system likely depends on higher-order or nonlinear correlations. Moreover, the non-selected feature VEC shows no discernible pattern, further emphasizing the complexity of phase discrimination in Heusler.

In challenging classification regimes, we leveraged quantum boosting primarily as a feature selection engine. The resulting sparse, interpretable feature sets guided follow-up modeling using QSVM, allowing for more expressive decision boundaries and deeper mechanistic insight. This flexible dual use, as both a classifier and a transparent feature selector, positions quantum boosting as a powerful and versatile component within quantum-enhanced machine learning pipelines for materials discovery.

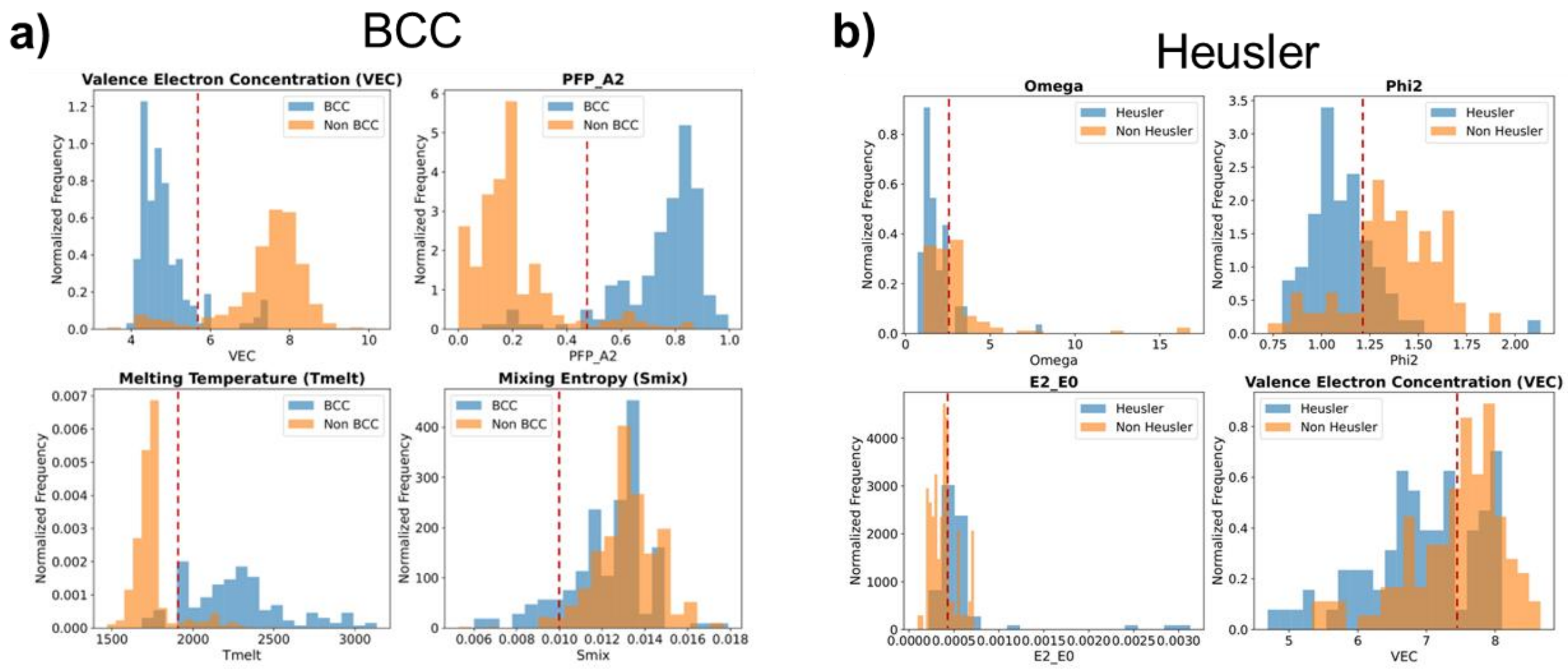

**Fig 4. Distributions of selected and non-selected features in phase vs. non-phase materials.** (a) Normalized histograms comparing BCC and non-BCC materials for four features: VEC, PFP_A2, Tmelt (selected by QBoost), and Smix (not selected). (b) Same for Heusler vs. non-Heusler, showing Omega, Phi2, and E2_E0 (selected), along with VEC (not selected). Vertical red dashed lines mark the decision thresholds identified by individual weak classifiers.

## Quantum Support Vector Machine

Building upon the interpretability and generalization strengths of quantum-enhanced methods, we implemented a Quantum Support Vector Machine model, leveraging features selected by Quantum Boosting. Inspired by the findings of Willsch et al. **[29]**, who demonstrated superior generalization of QSVMs under constrained regimes, we adapted their methodology for the task of HEA phase classification.

The QSVM algorithm is based on optimizing the dual loss function of a support vector machine, defined as:

$$L(\alpha) = \frac{1}{2}\sum_{n,m} \alpha_n \alpha_m y_n y_m K(x_n, x_m) - \sum_{n} \alpha_n, \qquad (3)$$

Subject to the constraints:

$$0 \leq \alpha_n \leq C, \qquad \sum\nolimits_n \alpha_n y_n = 0, \tag{4}$$

where $\alpha_n$ are the dual coefficients, $y_n \in \{-1, +1\}$ are the class labels, $K(\cdot, \cdot)$ the kernel function, and $C$ the box constraint parameter.

In the QSVM formulation, the dual coefficients $\alpha_n$ are discretized into a binary representation suitable for quantum annealing. Specifically, each coefficient $\alpha_n$ is represented as a binary expansion using $K$ bits and base $B$:

$$\alpha_n = \sum_{k=0}^{K-1} B^k a_{Kn+k}, \qquad a_{Kn+k} \in \{0,1\}, \tag{5}$$

Thus, implicitly define the regularization parameter $C$ as:

$$C = \sum_{k=0}^{K-1} B^k. \tag{6}$$

To ensure the equality constraint $\sum_n \alpha_n y_n = 0$ (essential for defining a well-posed decision boundary), we introduce a penalty term with multiplier $\xi$. The resulting QUBO formulation integrates both the radial basis function (RBF) kernel $K(x_n, x_m) = e^{-\gamma \|\vec{x}_n - \vec{x}_m\|^2}$ and constraint penalties, mapping the training objective onto an Ising energy landscape solvable via quantum annealing (additional details are available in the Methods section).

After quantum annealing, the recovered binary states are decoded into discrete dual coefficients $\alpha_n$, which determine the QSVM decision function. For direct comparison with classical SVMs (CSVM), we matched regularization constants via this encoding scheme and tuned the RBF kernel parameter $\gamma$ consistently across both quantum and classical implementations. Further details on the implementation and parameter tuning procedures are provided in the Methods section.

We benchmarked QSVM and CSVM using 5-fold cross-validation on our curated alloy database and assessed generalization on our independent test set of experimentally synthesized compounds, focusing on accuracy and F1-score. While overall performance was comparable, QSVM occasionally surpassed CSVM on the validation folds and consistently demonstrated stronger generalization to the experimental test set (see Table 3 for detailed metrics). This suggests that the stochastic solution landscape of the quantum annealer, combined with discretized support vector weights, may act as an implicit regularizer **[29]**.

To further validate our approach, we evaluated model performance on a curated test set of 86 HEAs that were experimentally synthesized and characterized in out labs as a part of a previous study [**20**]. These compositions were randomly drawn from element sets present in the training data, with the goal of spanning both within and beyond the original feature space. The distribution of the test alloys reflects the phase diversity of the training data, ensuring representative coverage across the modeled landscape. A comprehensive comparison between predicted and experimentally observed phases is provided in Table 4, offering an alloy-by-alloy breakdown across six structural phase categories. This evaluation highlights both the model's strengths and its limitations, particularly in borderline cases involving multiphase character or subtle intermetallic ordering.

**Table 3. Accuracy and F1-scores for QSVM vs. CSVM (per phase).** Accuracy and macro-averaged F1 scores are reported for QSVM and CSVM. Performance is evaluated on both training (5-fold cross-validation) and testing (unseen experimentally synthesized alloys). The parameters of the QA version of the SVM are $QSVM(B, K, \xi, \gamma)$ where $B$ is the encoding base, $K$ is the number of qubits per coefficient $\alpha_n$, $\xi$ is a Lagrangian multiplier, and $\gamma$ is the kernel parameter. The corresponding version of the classical SVM is $CSVM(C, \gamma)$.

| Phase | SVM Parameters | Validation | | Testing | |
|---|---|---|---|---|---|
| | | Accuracy | F1-score | Accuracy | F1-score |
| FCC | QSVM(5, 2, 0, 0.125) | 0.92 | 0.92 | 0.90 | 0.90 |
| | CSVM(6, 0.125) | 0.89 | 0.88 | **0.93** | 0.83 |
| BCC | QSVM(3, 3, 0, 0.125) | 0.94 | 0.94 | 0.92 | 0.92 |
| | CSVM(13, 0.125) | 0.93 | 0.92 | **0.95** | 0.89 |
| RB2 | QSVM(2, 2, 0, 0.125) | 0.80 | 0.78 | **0.93** | 0.93 |
| | CSVM(3, 0.125) | 0.69 | 0.73 | 0.86 | 0.89 |
| Heusler | QSVM(2, 3, 0, 0.125) | 0.85 | 0.85 | **0.83** | 0.83 |
| | CSVM(7, 0.125) | 0.82 | 0.78 | 0.75 | 0.73 |
| Laves | QSVM(5, 2, 0, 0.25) | 0.91 | 0.91 | **0.90** | 0.91 |
| | CSVM(6, 0.25) | 0.87 | 0.85 | 0.90 | 0.7 |
| Sigma | QSVM(5, 2, 0, 0.125) | 0.87 | 0.86 | **0.95** | 0.95 |
| | CSVM(6, 0.125) | 0.83 | 0.79 | 0.80 | 0.40 |

A deeper analysis revealed that CSVM tended to excel in cleanly separable cases such as FCC and BCC. In contrast, QSVM gained an edge in more challenging regimes, including complex intermetallic phases, where sub-optimal but diverse solutions could better capture nonlinear structure and suppress overfitting.

To prove this further, we evaluated the CSVM dual loss using both classical and quantum-derived support vector coefficients (Fig. 5). As expected, CSVM always achieved the lowest loss, consistent with convexity and deterministic solvers. QSVM solutions approached, but did not undercut, this minimum, reflecting the annealer's tendency to return near-optimal, rather than globally optimal, configurations**[29]**.

Interestingly, the best-performing QSVM models emerged when the constraint $\sum_n \alpha_n y_n = 0$ was relaxed rather than strictly enforced. This condition, which originates from the Karush-Kuhn-Tucker (KKT) conditions in the dual SVM formulation, ensures the existence of a well-defined bias term and optimal margin placement**[47]**. At a high penalty value ($\xi = 5$), the constraint term dominated the objective function, severely distorting the evaluated loss (Fig. 5). In contrast, $\xi = 0$ consistently yielded superior generalization performance, as evidenced in Table 3, where the best-performing models always corresponded to $\xi = 0$.

These observations align with principles from robust optimization and margin theory**[48,49]**. Strict enforcement of the equality constraint can limit the expressiveness of the support vector set, effectively narrowing the solution landscape**[50, 51]**. In contrast, soft constraints allow the quantum annealer to explore a richer set of near-optimal models that, while not globally minimizing the dual loss, offer improved robustness to overfitting, analogous to early stopping in neural networks or regularization via noisy optimization paths**[52,53]**.

Furthermore, this behavior parallels observations from Willsch et al. **[29]** and He & Xiao **[34]**, who noted that the stochastic nature of quantum annealing introduces implicit regularization. We extend this by demonstrating that relaxing constraint compliance explicitly contributes to this effect.

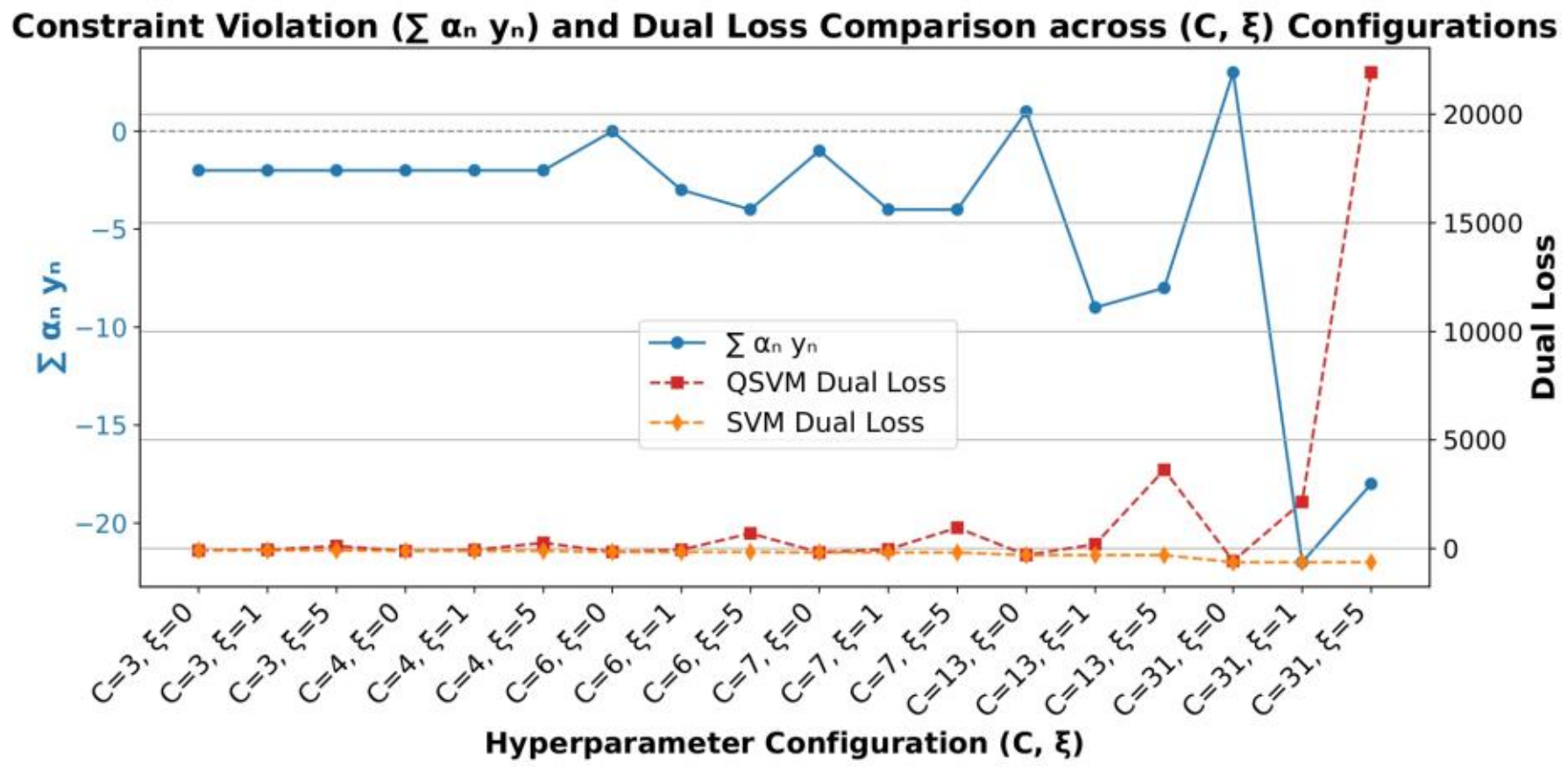

**Figure 5. QSVM dual-loss and constraint violation analysis across hyperparameter settings.** QSVM dual-loss (red squares) and classical SVM dual-loss (orange diamonds) compared with constraint violations (blue circles) of QSVM for RB2 dataset.

This behavior parallels phenomena in neural network training, where methods such as early stopping or allowing margin violations often lead to improved generalization compared to strict optimization**[52, 53]**. In this light, QSVM can be viewed not simply as an approximate SVM, but as a stochastic regularized variant, one that trades strict KKT compliance for broader exploration of low-energy, high-performing hypotheses.

To illustrate this visually, we selected the RB2 phase dataset, where QSVM showed the largest difference between validation and test accuracy. Using only two interpretable features, Smix and Omega, we constructed binary classifiers using both SVM and QSVM. We then plotted the decision functions alongside the training and testing data distributions in a 2×2 grid (Fig. 6). Despite the CSVM solution yielding a marginally lower loss, the QSVM decision boundary captured the separation more robustly on the test data, especially in ambiguous regions near the classification margin.

These results emphasize that, while classical optimization retains its advantage in precise loss minimization, quantum-enhanced classifiers can offer tangible benefits in generalization. The constrained, ensemble-driven search carried out by quantum annealing appears well-suited for classification tasks involving physical descriptors, especially in regimes where data is scarce, features are correlated, or relationships are nonlinear.

Our findings collectively underscore the significant potential for quantum-enhanced classifiers to not only improve accuracy but also deliver physically intuitive, robust predictive models. Future efforts should aim at scaling this methodology further, exploring more efficient encodings, and leveraging upcoming hardware features, such as reverse annealing or advanced embedding heuristics, to further improve training fidelity and model generalization.

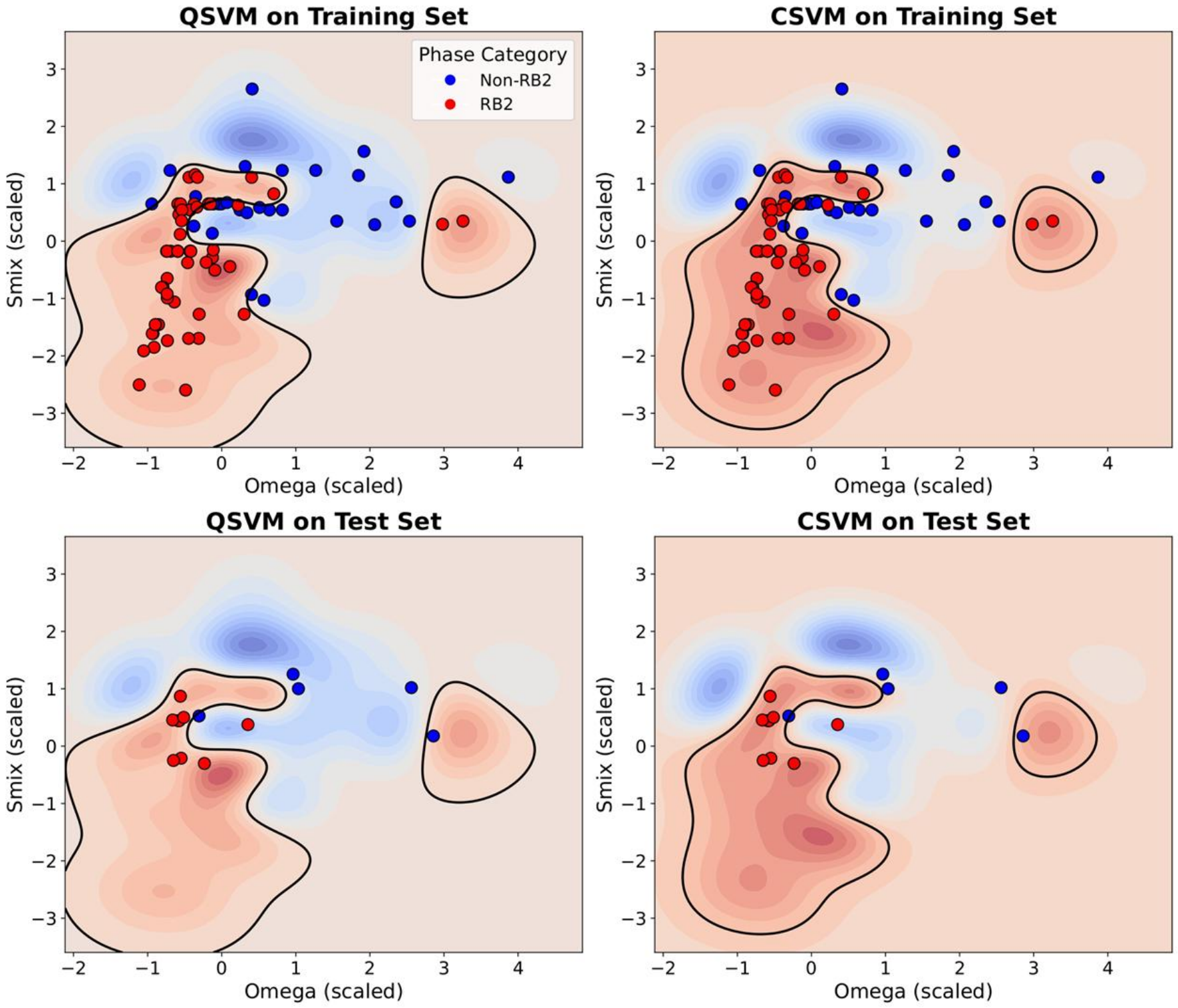

**Figure 6. QSVM and CSVM decision boundaries for RB2 phase.** QSVM (left) and CSVM (right) show training set performance (top) and generalization to test set (bottom), based on two interpretable features: Smix and Omega. QSVM better captures nonlinear boundaries on unseen data, aligning more closely with the true phase distribution.

**Table 4. Testing HEAs compositions, phase prediction results from QSVM, and experimental phase characterization results obtained from XRD are listed.** This table lists the 86 High-Entropy Alloy compositions used for final testing, along with the corresponding phase predictions and experimental results obtained from X-ray diffraction (XRD). Phase predictions are given as "True" or "False". Experimental results detail the observed phase assemblages, using the following abbreviations: A1 (FCC), A2 (BCC), Mix A1–A2 (coexisting A1/A2 or their mixture), AlNi B2+ (AlNi-type B2 phase coexisting with other phases), Al-X–Y B2+ (refractory-type B2), and $L2_1$ (Heusler). Incorrect predictions are highlighted in bold and underlined.

A. FCC, BCC, Laves+, and Sigma+ prediction models validation HEAs

| Composition | FCC Prediction | BCC Prediction | Laves+ Prediction | Sigma+ Prediction | Experimental results |
|---|---|---|---|---|---|
| $Ag_{20}Al_{20}Cr_{20}Mn_{20}Ni_{20}$ | FALSE | FALSE | FALSE | FALSE | B2+A1 |
| $Ag_{5}Al_{38}Cr_{19}Mn_{19}Ni_{19}$ | FALSE | FALSE | FALSE | FALSE | B2+A1 |
| $Al_{5}Co_{20}Cr_{10}Fe_{40}Ni_{20}Ti_{5}$ | TRUE | FALSE | FALSE | FALSE | A1 |
| $Al_{10}Co_{20}Cu_{20}Fe_{20}Ni_{20}V_{10}$ | FALSE | FALSE | FALSE | FALSE | B2+A1 |
| $Al_{11}Co_{22}Cr_{11}Cu_{11}Ni_{33}V_{12}$ | FALSE | FALSE | FALSE | FALSE | B2+A1 |
| $Al_{15}Cr_{15}Mo_{15}Ni_{46}W_{9}$ | FALSE | FALSE | FALSE | FALSE | B2+A1+A2 |
| $Al_{15}Cr_{31}Fe_{31}Mn_{15}Ni_{8}$ | FALSE | FALSE | FALSE | FALSE | B2 |
| $Al_{16}Co_{20}Fe_{20}Mn_{18}Ni_{20}V_{6}$ | FALSE | FALSE | FALSE | FALSE | B2 |
| $Al_{16}Co_{21}Cr_{21}Fe_{21}Ni_{21}$ | FALSE | FALSE | FALSE | FALSE | B2+A1 |
| $Al_{16}Cr_{16}Fe_{16}Mn_{16}Ni_{31}V_{5}$ | FALSE | FALSE | FALSE | FALSE | B2 |
| $Al_{19}Cr_{19}Cu_{19}Fe_{19}Ni_{19}Si_{5}$ | FALSE | FALSE | FALSE | FALSE | B2+A1+A2 |
| $Al_{20}Co_{20}Cr_{20}Fe_{20}Mn_{20}$ | FALSE | FALSE | FALSE | FALSE | B2 |
| $Al_{21}Co_{11}Cr_{21}Cu_{5}Fe_{21}Mn_{21}$ | FALSE | FALSE | FALSE | FALSE | B2 |
| $Al_{22}Co_{26}Fe_{26}Ni_{26}$ | FALSE | **FALSE** | FALSE | FALSE | A2 |
| $Al_{23}Co_{23}Cu_{23}Fe_{23}V_{8}$ | FALSE | FALSE | FALSE | FALSE | B2+A1 |
| $Al_{23}Cu_{23}Fe_{23}Ni_{23}V_{8}$ | FALSE | FALSE | FALSE | FALSE | B2+A1 |
| $Al_{24}Co_{24}Cu_{23}Ni_{23}Ti_{6}$ | FALSE | FALSE | FALSE | FALSE | B2+A1 |
| $Al_{25}Co_{25}Cr_{25}Fe_{25}$ | FALSE | FALSE | FALSE | FALSE | B2 |
| $Al_{25}Cu_{25}Fe_{25}Ni_{25}$ | FALSE | FALSE | FALSE | FALSE | B2+A1 |
| $Al_{29}Co_{29}Cu_{13}Fe_{29}$ | FALSE | FALSE | FALSE | FALSE | B2+A1 |
| $Al_{33}Co_{17}Nb_{33}Ni_{17}$ | FALSE | **TRUE** | FALSE | FALSE | B2+Laves |
| $Co_{7}Ta_{31}Ti_{31}V_{31}$ | FALSE | TRUE | **TRUE** | FALSE | A2 |
| $Cr_{6}Ti_{56}V_{19}Zr_{19}$ | FALSE | TRUE | **TRUE** | FALSE | A2 |
| $Cr_{25}Mo_{25}Ti_{25}V_{25}$ | FALSE | TRUE | FALSE | FALSE | A2 |
| $Cr_{33}Mo_{22}Nb_{12}V_{33}$ | FALSE | TRUE | **TRUE** | FALSE | A2 |
| $Hf_{25}Nb_{25}Ta_{25}Zr_{25}$ | FALSE | TRUE | FALSE | FALSE | A2 |
| $Hf_{30}Nb_{30}Ti_{30}V_{10}$ | FALSE | TRUE | FALSE | FALSE | A2 |
| $Hf_{30}Ta_{30}Ti_{30}V_{10}$ | FALSE | TRUE | FALSE | FALSE | A2 |
| $Mo_{29}Nb_{13}Ti_{29}V_{29}$ | FALSE | TRUE | FALSE | FALSE | A2 |
| $Nb_{22}Ta_{22}Ti_{22}V_{22}Zr_{12}$ | FALSE | TRUE | FALSE | FALSE | A2 |
| $Nb_{29}Ta_{29}Ti_{29}Zr_{13}$ | FALSE | TRUE | FALSE | FALSE | A2 |
| $Co_{15}Cr_{15}Fe_{15}Mn_{15}Ni_{32}V_{8}$ | TRUE | FALSE | FALSE | **TRUE** | A1 |
| $Co_{18}Cu_{18}Fe_{18}Mn_{18}Ni_{18}V_{10}$ | **FALSE** | FALSE | FALSE | FALSE | A1+A1 |
| $Co_{19}Cr_{29}Fe_{29}Ni_{19}Si_{4}$ | TRUE | FALSE | FALSE | FALSE | A1 |
| $Co_{21}Cr_{11}Fe_{42}Ni_{21}Ti_{5}$ | TRUE | FALSE | **TRUE** | FALSE | A1 |
| $Co_{22}Fe_{22}Mn_{12}Ni_{44}$ | TRUE | FALSE | FALSE | FALSE | A1 |
| $Co_{24}Cr_{24}Fe_{24}Ni_{24}Si_{4}$ | TRUE | FALSE | FALSE | FALSE | A1 |
| $Co_{24}Fe_{24}Ni_{47}V_{5}$ | TRUE | FALSE | FALSE | FALSE | A1 |
| $Co_{25}Cr_{8}Cu_{5}Fe_{25}Ni_{25}V_{12}$ | TRUE | FALSE | FALSE | FALSE | A1 |
| $Cr_{19}Cu_{19}Fe_{19}Mn_{18}Ni_{19}Ti_{6}$ | **TRUE** | FALSE | FALSE | FALSE | A1+A2 |
| $Al_{4}Cr_{32}Cu_{32}Fe_{11}Mn_{21}$ | FALSE | FALSE | FALSE | FALSE | A1+A2 |
| $Al_{8}Cr_{56}Fe_{14}Mn_{22}$ | FALSE | TRUE | FALSE | FALSE | A2 |
| $Al_{10}Co_{20}Cr_{10}Cu_{20}Mn_{20}Ni_{20}$ | TRUE | FALSE | FALSE | FALSE | A1+A2 |
| $Al_{24}Co_{24}Cr_{23}Fe_{23}Ti_{6}$ | FALSE | FALSE | FALSE | FALSE | B2 |
| $Co_{16}Cr_{16}Cu_{16}Fe_{16}Mn_{14}Ni_{16}Ti_{6}$ | TRUE | FALSE | FALSE | FALSE | A1+A1 |
| $Co_{25}Cr_{25}Cu_{25}Fe_{25}$ | FALSE | FALSE | FALSE | FALSE | A1+A1+Unknown |
| $Cr_{25}Cu_{25}Fe_{25}Mn_{25}$ | FALSE | FALSE | FALSE | FALSE | A1+A2 |
| $Cr_{40}Fe_{40}Mn_{10}Ni_{10}$ | FALSE | TRUE | FALSE | FALSE | A2 |
| $Co_{20}Fe_{20}Mn_{20}Ni_{20}Ti_{10}V_{10}$ | FALSE | FALSE | TRUE | FALSE | Laves+A2 |
| $Co_{20}Fe_{20}Mo_{20}Ni_{20}Ti_{20}$ | FALSE | FALSE | TRUE | FALSE | Laves+A1+A2 |
| $Co_{21}Cr_{21}Cu_{21}Mn_{16}Ti_{21}$ | FALSE | **TRUE** | TRUE | FALSE | Laves+A1 |
| $Co_{25}Cr_{25}Fe_{25}Nb_{13}Ti_{12}$ | FALSE | FALSE | TRUE | FALSE | Laves+A1+A2 |
| $Cr_{20}Nb_{20}Ni_{20}Ti_{20}Zr_{20}$ | FALSE | **TRUE** | TRUE | FALSE | Laves+A2 |
| $Cr_{40}Fe_{20}Ni_{20}Ti_{20}$ | FALSE | FALSE | TRUE | FALSE | Laves+A1+A2 |
| $Cu_{17}Fe_{17}Mn_{17}Ni_{17}Ti_{32}$ | **TRUE** | **TRUE** | TRUE | FALSE | Laves+A1+A2 |
| $Co_{15}Cr_{15}Cu_{8}Fe_{15}Ni_{31}Ti_{8}V_{8}$ | TRUE | FALSE | **TRUE** | FALSE | A1 |
| $Co_{18}Cr_{18}Fe_{18}Mo_{18}Ni_{18}V_{10}$ | FALSE | FALSE | FALSE | TRUE | Sigma+A1 |
| $Co_{20}Cr_{20}Fe_{20}Mo_{20}V_{20}$ | FALSE | FALSE | FALSE | TRUE | Sigma+A2 |
| $Co_{26}Cr_{26}Fe_{26}Mo_{22}$ | **TRUE** | FALSE | FALSE | **FALSE** | Sigma+A2 |
| $Cu_{20}Fe_{20}Mn_{20}Ni_{20}V_{20}$ | FALSE | FALSE | FALSE | **FALSE** | Sigma+A1 |

B. Al-X-Y type B2+ prediction model validation HEAs

| Composition | Al-X-Y B2 + prediction | Experimental results | | Composition | Al-X-Y B2 + prediction | Experimental results |
|---|---|---|---|---|---|---|
| $Al_{10}Hf_{20}Nb_{22}Ti_{33}V_{15}$ | TRUE | B2 | | $Al_{30}Nb_{20}Ta_{15}Ti_{20}V_{10}Zr_{5}$ | TRUE | B2 |
| $Al_{15}Hf_{25}Nb_{32}Ti_{28}$ | TRUE | B2 | | $Al_{30}Nb_{20}Ta_{20}Ti_{20}Zr_{10}$ | TRUE | B2+Unknown |
| $Al_{20}Hf_{24}Nb_{29}Ti_{27}$ | TRUE | B2 | | $Al_{4}Hf_{6}Nb_{42}Ti_{18}V_{24}W_{6}$ | FALSE | A2 |
| $Al_{23}Hf_{23}Nb_{23}Ti_{23}V_{8}$ | TRUE | B2 | | $Al_{8}Cr_{15}Mo_{15}Nb_{15}Ti_{15}V_{32}$ | FALSE | A2 |
| $Al_{23}Hf_{23}Ta_{23}Ti_{23}V_{8}$ | TRUE | B2 | | $Al_{10}Hf_{18}Nb_{18}Ta_{18}Ti_{18}Zr_{18}$ | FALSE | A2 |
| $Al_{26}Mo_{21}Nb_{11}Ti_{21}V_{21}$ | **TRUE** | A2 | | $Al_{32}Nb_{17}Ta_{17}Ti_{17}V_{17}$ | TRUE | B2 |
| $Al_{30}Mo_{20}Nb_{20}Ti_{30}$ | TRUE | B2 | | $Al_{6}Nb_{21}Ta_{21}Ti_{21}V_{21}Zr_{10}$ | FALSE | A2 |

C. Heusler+ prediction model validation HEAs

| Composition | Heusler+ prediction | Experimental results | | Composition | Heusler+ prediction | Experimental results |
|---|---|---|---|---|---|---|
| $Al_{10}Co_{25}Fe_{25}Mn_{25}Ti_{15}$ | FALSE | A2+Unknown | | $Al_{25}Cr_{10}Fe_{20}Mn_{10}Ni_{20}Ti_{15}$ | **FALSE** | $L2_1$ |
| $Al_{10}Cr_{5}Fe_{45}Mn_{12}Ni_{20}Ti_{8}$ | TRUE | $L2_1$+A1 | | $Al_{10}Co_{20}Mn_{20}Ni_{30}Ti_{10}$ | FALSE | A1+A2 |
| $Al_{12}Co_{28}Fe_{19}Ni_{29}Ti_{12}$ | **FALSE** | $L2_1$+A1 | | $Al_{10}Co_{30}Fe_{20}Ni_{32}Ti_{8}$ | FALSE | A1 |
| $Al_{14}Cr_{4}Fe_{17}Mn_{4}Mo_{1}Ni_{44}Ti_{16}$ | TRUE | $L2_1$+A1 | | $Al_{15}Co_{30}Fe_{30}Ni_{10}Ti_{15}$ | FALSE | B2 |
| $Al_{15}Cr_{10}Fe_{30}Ni_{30}Ti_{15}$ | TRUE | $L2_1$+A2 | | $Al_{20}Fe_{10}Mn_{30}Ni_{40}$ | FALSE | B2+A1 |
| $Al_{15}Fe_{40}Mn_{20}Ni_{10}Ti_{10}$ | TRUE | $L2_1$ | | $Al_{7}Co_{30}Fe_{30}Mn_{25}Ti_{8}$ | FALSE | B2+A1 |

## Discussion

Our exploration of quantum-enhanced machine learning techniques, namely Quantum Boosting (QBoost) and Quantum Support Vector Machines (QSVM), has demonstrated notable advantages in predicting crystallographic phases of High Entropy Alloys (HEAs). A central finding of this work is the beneficial impact of quantum optimization, particularly when paired with physically interpretable feature sets. Our quantum-enhanced framework consistently delivers competitive, and often superior, performance relative to classical approaches **[20]**, with the critical added benefits of enhanced interpretability and reduced computational cost .

One promising direction for further improving QBoost lies in the introduction of mathematical operations between input features prior to assignment to weak classifiers. Allowing the use of derived descriptors, such as ratios or nonlinear combinations of thermodynamic variables, may improve accuracy by better capturing complex, physically meaningful relationships **[20]**. However, while expanding expressiveness, such transformations must be carefully designed to preserve interpretability. Excessive complexity could undermine the principle of "weakness" in the base learners, diluting the inherent clarity that QBoost offers.

Additionally, QBoost's binary weighting scheme, which enforces sparsity and aids interpretability, could be extended to include continuous weights, akin to our encoding strategy in QSVM. While this could enhance model flexibility, it introduces new optimization challenges and risks drifting toward classical boosting models like AdaBoost. A middle ground, incorporating moderately fine-grained weight encodings, may strike a useful balance between simplicity and performance, depending on application needs.

Our QSVM approach achieves improved performance largely due to a novel formulation of the loss function for compatibility with Quadratic Unconstrained Binary Optimization (QUBO). Unlike prior implementations that emphasized quantum fluctuation effects, our approach introduces a relaxed constraint on the dual coefficients $\sum_n \alpha_n y_n$, as suggested by Willsch et al. **[29]**. Empirically, we find that allowing minor violations of this condition, moderated by a penalty parameter, yields a richer ensemble of near-optimal solutions. These higher-energy quantum solutions, while not strictly optimal under the convex classical formulation, consistently demonstrate better generalization, likely due to their ability to capture nonlinear patterns without overfitting.

When benchmarked against our previous intermetallic phase prediction model **[20]**, which relied on extensive feature engineering and sequential selection, our quantum-enhanced pipeline yielded notable performance gains, particularly for the Laves, Sigma and RB2 phases. RB2 prediction accuracy improved from 86% to 93%, despite using a leaner and less computationally intensive feature set. Conversely, performance on Heusler phases still favored the previous method, indicating that this phase may require further model refinement or targeted hybrid strategies to capture their inherent structural complexity.

A current limitation of our QSVM implementation arises from the quantized nature of the regularization parameter $C$, which is tied to the binary encoding of support vector coefficients. This restriction reduces tunability and may cap performance in certain cases. However, we also observe that this discretization, combined with quantum annealing's stochastic search behavior, acts as an implicit regularizer, especially in nonlinear, data-sparse regimes like RB2 or Sigma. This suggests that the QSVM's advantage may lie less in precise optimization and more in its ability to explore diverse low-energy solutions, promoting generalization through controlled constraint violation and noise-driven sampling **[29]**.

Importantly, all models were evaluated not only through cross-validation, but also against an independent test set of HEAs that we experimentally synthesized and characterized via X-ray diffraction on a prior study **[20]** (see Supplementary Fig. 7). These alloys were entirely excluded from training or hyperparameter tuning, providing a rigorous, real-world benchmark of generalization. The strong alignment between model predictions and experimentally measured phases reinforces the practical value of this quantum-enhanced approach and demonstrates the feasibility of integrating these methods into real-world materials discovery pipelines.

While the speedups we observe for quantum annealing over simulated annealing (SA) are substantial, reaching up to four orders of magnitude, these results are inherently hardware-dependent. Prior work has shown that, under certain classical computing architectures or with highly optimized SA implementations, classical methods can approach the performance of quantum annealers **[54-56].** Nevertheless, platforms like D-Wave offer accessible cloud-based quantum annealing resources, potentially lowering the barrier to entry compared to the specialized HPC infrastructure needed for large-scale classical SA. As quantum hardware continues to evolve, future benchmarking efforts should aim to compare runtime and scalability across diverse hardware platforms and problem formulations, ensuring robust and reproducible performance assessments.

In summary, our results establish that quantum-enhanced machine learning, particularly through interpretable models like Quantum Boosting and regularized classifiers like Quantum Support Vector Machines, offers a compelling new toolset for materials informatics. Future directions include deeper feature transformations, more flexible encoding schemes, and hybrid inference strategies to further improve accuracy on structurally complex phases like Heusler. With increasing accessibility of quantum resources and a growing library of quantum-amenable algorithms, this work has progressed a step toward practical, scalable quantum acceleration in the design of advanced materials.

## METHODS

### Experimental Methods

The 86 high-entropy alloy compositions used as our independent test set were experimentally synthesized and characterized in a previous study by our group **[20].** High- purity raw elements (>99.99 wt. %) were arc melted in a water-cooled copper crucible using a 400-A current. Each melt lasted 30s, with at least five melts performed to ensure chemical homogeneity, and the ingots were flipped between each melt. The ingots were later polished for XRD analysis down to a grit size of 1μm with diamond suspension and finished with a 0.06-μm colloidal silica suspension. The XRD measurements were conducted on the polished sample by a PANalytical Empyrean diffractometer with Cu Kα radiation and a scanning rate of approximately 0.15 degrees/s. XRD patterns can be found on Supplementary Fig. 7.

These previously synthesized and measured alloys were not used during training or hyperparameter tuning. Instead, they serve as a rigorously curated, real-world test set to evaluate the generalization performance of our quantum-classical models.

### Data Construction

To enable robust machine learning (ML) prediction of phase formation in high-entropy alloys (HEAs), we curated separate datasets for six targeted phase types: FCC, BCC, Sigma , Laves, Heusler, and refractory Al-X–Y type B2 phases. Each dataset was constructed to capture both presence and absence of the phase of interest, with careful attention to class balance and compositional validity.

For FCC phase, the dataset comprises 132 HEAs with the FCC phase (Laves) and 698 without (No-FCC) Due to significant class imbalance, we employed random under-sampling of the majority class to construct balanced subsets (n = 132 per class). This process was repeated over 30 independent sampling rounds to ensure robustness. This process was used for imbalanced classes such as BCC, Sigma, and Laves.

For BCC phase, the dataset comprises 178 HEAs with the FCC phase (Laves) and 652 without (No-FCC). Under-sampling was used to generate balanced training sets across 30 iterations.

For the Sigma phase, 52 HEAs containing σ-phase (Sigma) and 783 without it (No-Sigma) were identified. Under-sampling was used to generate balanced training sets across 30 iterations.

The Laves phase dataset comprises 96 HEAs with the phase (Laves) and 739 without (No-Laves). Under-sampling was used to generate balanced training sets across 30 iterations.

For the Heusler phase, characterized by the $X_2YZ$ stoichiometry, 77 HEAs exhibiting the phase ($L2_1$) were collected. The 109 HEAs without the phase (Non-$L2_1$) were stringently selected to include valid X, Y, and Z elements.

The Al-X–Y type B2 phase dataset focuses on refractory systems containing aluminum and combinations of X = {Ti, Zr, Hf} and Y = {Cr, Mo, Nb, V}. Here, 52 HEAs with the B2 phase (Al-X-Y B2) and 35 compositionally similar HEAs without it (Non- Al-X-Y B2) were identified. All samples meet the necessary stoichiometric constraints for phase formation.

Full datasets and descriptors are provided in the Supplementary Information.

## Classical Models

Baseline models including Support Vector Machines (SVM) and decision trees were implemented using the scikit-learn library. Hyperparameters were tuned via grid search using 5-fold cross-validation.

For SVM models, all input features were standardized to zero mean and unit variance. No feature scaling was applied for boosting models, as decision stumps are scale-invariant and rely on the natural interpretability of raw physical descriptors.

## Quantum/Classical Annealing: Hardware and Implementation

Quantum annealing was performed using the D-Wave Advantage system via the Ocean SDK on the Pegasus topology. For Quantum Boosting (QBoost), minor embeddings were generated using the default heuristic algorithm. Logical qubits were represented as chains of physical qubits with a chain strength of 2.0, empirically optimized to minimize chain breaks. Each QUBO instance was sampled 1,000 times with an annealing time of 100μs; the lowest-energy solution was retained.

Simulated Annealing (SA) baselines were executed using D-Wave's reference implementation from Ocean SDK, using dimod package, matching the QA sample count and using default parameters. All SA runs were performed on UVA's Rivanna High-Performance Computing cluster equipped with Intel Xeon Platinum CPUs.

Quantum Support Vector Machine (QSVM) optimization was performed using D-Wave's Leap Hybrid Solver due to the dense kernel matrices and embedding challenges.

## Quantum Boosting

Each weak classifier consisted of a decision stump (depth-1 decision tree) trained on a single unscaled feature. A total of 31 classifiers, one per feature, were used. Cross-validation (5-fold) was subsequently performed on the trained model to evaluate accuracy and F1-score. Hyperparameter tuning was conducted over $\lambda \in \{1, 0.1, 0.01, 0.001, 0.001\}$ followed by localized refinement around the best-performing value (as selected by CV accuracy), with generalization assessed on a held-out test set.

To construct the QUBO we followed the same steps as in **[30, 31]**

The boosting loss function minimized over binary weights $\omega_i \in \{0,1\}$ was:

$$\min_{\omega \in \{0,1\}^N} \left[ \sum_{s=1}^{S} \left( y_s - \sum_{i=1}^{N} \omega_i h_i(x_s) \right)^2 + \lambda \|\omega\|_0 \right]$$

This expands to a QUBO form:

$$\min_{\omega \in \{0,1\}^N} \left[ \sum_{i,j=1}^{N} C'_{ij} \omega_i \omega_j + 2 \sum_{i=1}^{N} \left( \lambda - C'_{iy} \right) \omega_i \right]$$

where $C'_{ij} = \sum_s h_i(x_s) h_j(x_s)$ and $C'_{iy} = \sum_s h_i(x_s) y_s$.

To convert to an Ising model, binary variables were transformed to spin variables via $q_i = 2\omega_i - 1$, where $q_i \in \{-1, 1\}$. This yields the final Ising Hamiltonian:

$$H_F = \sum_{i,j=1}^{N} C_{ij} q_i q_j + 2 \sum_{i=1}^{N} (\lambda - C_{iy}) q_i$$

with $C_{ij} = \frac{1}{4} C'_{ij}$, and $C_{iy} = C'_{iy} - \frac{1}{2} \sum_j C'_{ij}$ . The resulting Hamiltonian was compiled and embedded for execution on D-Wave's quantum annealing hardware. Detail derivation can be found on **[30, 31].**

## Quantum Support Vector Machine

The classical dual loss function for the Support Vector Machine (SVM) formulation we address is given by:

$$L(\alpha) = \frac{1}{2} \sum_{n,m} \alpha_n \alpha_m y_n y_m K(x_n, x_m) - \sum_n \alpha_n,$$

where $\alpha_n$ are the dual coefficients, $y_n \in \{-1, +1\}$ are the binary class labels, $K(x_n, x_m)$ is the kernel function (here an RBF kernel $K(x_n, x_m) = e^{-\gamma \|\vec{x}_n - \vec{x}_m\|^2}$). This optimization is performed under the constraints:

$$0 \le \alpha_n \le C, \qquad \sum_n \alpha_n y_n = 0$$

Where $C$ is the box constraint regularization parameter that controls the trade-off between margin maximization and misclassification penalties.

The resulting classifier is defined by the decision function:

$$f(x) = \sum_n \alpha_n y_n K(x_n, x) + b,$$

where $b$ is the bias

$$b = \frac{\sum_n \alpha_n (C - \alpha_n) [y_n - \sum_m \alpha_m y_m K(x_m, x_n)]}{\sum_n \alpha_n (C - \alpha_n)}$$

computed using the support vectors. This formulation guarantees a global minimum due to the convexity of the loss function.

In contrast, the QSVM reformulates this convex optimization into a Quadratic Unconstrained Binary Optimization (QUBO) problem, which is compatible with quantum annealing hardware. To achieve this, each dual coefficient $\alpha_n$ is encoded as a binary expansion:

$$\alpha_n = \sum_{k=0}^{K-1} B^k a_{Kn+k}, \qquad a_{Kn+k} \in \{0,1\},$$

where $B$ is the encoding base and $K$ is the number of binary variables used to represent each coefficient. The maximum possible value defines the regularization constant $C$ implicitly as:

$$C = \sum_{k=0}^{K-1} B^k = \frac{B^k - 1}{B - 1}$$

To enforce the equality constraint $\sum_n \alpha_n y_n = 0$ which ensures the existence of a well-defined bias term in the SVM formulation, a penalty term is introduced with multiplier $\xi$, yielding the QUBO loss function:

$$E_{QUBO}(\alpha) = \frac{1}{2} \sum_{n,m,k,j} a_{Kn+k} a_{Km+j} B^{k+j} y_n y_m K(x_n, x_m) - \sum_{nk} B^k a_{Kn+k} + \xi \left( \sum_{nk} B^k a_{Kn+k} y_n \right)^2$$

$$E_{QUBO}(\alpha) = \frac{1}{2} \sum_{n,m,k,j} a_{Kn+k} Q_{Kn+k,Km+j} a_{Km+j},$$

where $Q$ is the binary-encoded QUBO matrix incorporating both the kernel interactions and the penalty term:

$$Q_{Kn+k,Km+j} = B^k B^j y_n y_m [K(x_n, x_m) + \xi] - \delta_{nm} \delta_{kj} B^k.$$

This formulation naturally maps onto the Ising or QUBO architecture supported by D-Wave systems.

After obtaining a solution from the hybrid solver, we reconstructed the discretized $\alpha_n$ values and evaluated the resulting QSVM decision function as 5-CV and later on the test data. To ensure fair comparison, we matched classical SVM and QSVM runs under identical values of $C$, dictated solely by the QUBO encoding, and performed hyperparameter tuning over the RBF kernel parameter $\gamma$.

Hyperparameter tuning was conducted over $B \in \{2, 3, 4\}$, $K \in \{2, 3\}$, $\xi \in \{0, 1, 5\}$, $\gamma = \in \{0.125, 0.25, 0.5, 1, 2, 4, 8\}$.

## ACKNOWLEDGMENTS

This work is supported by the Office of Naval Research under Grant No. N00014-23-1-2441.

## AUTHOR CONTRIBUTIONS

Conceptualization: D.I.H., G.-W.C., I.K., J.P., Methodology: D.I.H., Investigation: D.I.H., Writing-Original Draft: D.I.H., Writing – Review & Editing: G.-W.C., I.K., J.P., Supervision: G.-W.C., I.K., J.P.

## COMPETING INTERESTS

The authors declare no competing interests.

## MATERIALS & CORRESPONDENCE

Correspondence and requests for materials should be addressed to Diego Ibarra Hoyos (di8pd@virginia.edu) or Joseph Poon (sjp9x@virginia.edu).